\documentclass[%
reprint,
superscriptaddress,
 amsmath,amssymb,
 aps,
prl,
]{revtex4-2}

\usepackage{graphicx}
\usepackage{dcolumn}
\usepackage{bm}
\usepackage{float}

\begin{document}


\title{Conductance switching and nonequilibrium phase coexistence in superconductors with intermediate bias}

\author{Shamashis Sengupta}
\email[]{shamashis.sengupta@ijclab.in2p3.fr}
\affiliation{Universit\'{e} Paris-Saclay, CNRS/IN2P3, IJCLab, 91405 Orsay, France}

\begin{abstract}
Superconducting systems may display different types of nonequilibrium states depending on the specific constraints imposed for measurement. We probe current-voltage relations of three-dimensional superconducting films by allowing finite voltages to develop across their length. Our experiments reveal sharp features of negative differential conductance which highlight the validity of the principle of minimum entropy production at the critical current transition. We have observed dissipative states with resistances intermediate between those of superconducting and normal phases at zero applied magnetic field, indicating a phenomenon of phase coexistence under nonequilibrium conditions. The features of steady states reported here are not accessible in conventional transport experiments with current-biasing methods.
\end{abstract}

\maketitle

\newpage




\newpage

The physical properties of superconductors have proved to be extremely useful for conceiving novel devices in many different areas of research, like transition edge sensors\cite{clarke1,irwin}, spintronics\cite{linder,eschrig}, switches\cite{ritter,melnikov,mccaughan} and quantum computing technologies\cite{koch,barends,nakamura,clarke2}. The importance of superconductors is related to the fact that these systems can exist in two states with profoundly different electrical properties, allowing various means of manipulating them according to the principles of classical and quantum physics. A key feature of a superconductor is its ability to conduct electricity with exactly zero resistance, which implies that imposing an electric field must lead to a breakdown of superconducting order. Exploring how the system responds to such constraints is the main objective of this work. The problem of conduction in linear resistors can be described using the criterion of minimum dissipation\cite{landauer}. However, a superconductor presents a much more complex case of a non-linear circuit element that can behave differently under conditions of biasing with a current or a voltage. We describe here the method of intermediate biasing which allows us to observe steady states with less restrictive boundary conditions. Our measurements reveal a very sharp feature of negative differential conductance and steady states bearing signatures of coexisting phases that are not accessible in conventional current-biased experiments.

\begin{figure*}
\begin{center}
\includegraphics[width=180mm]{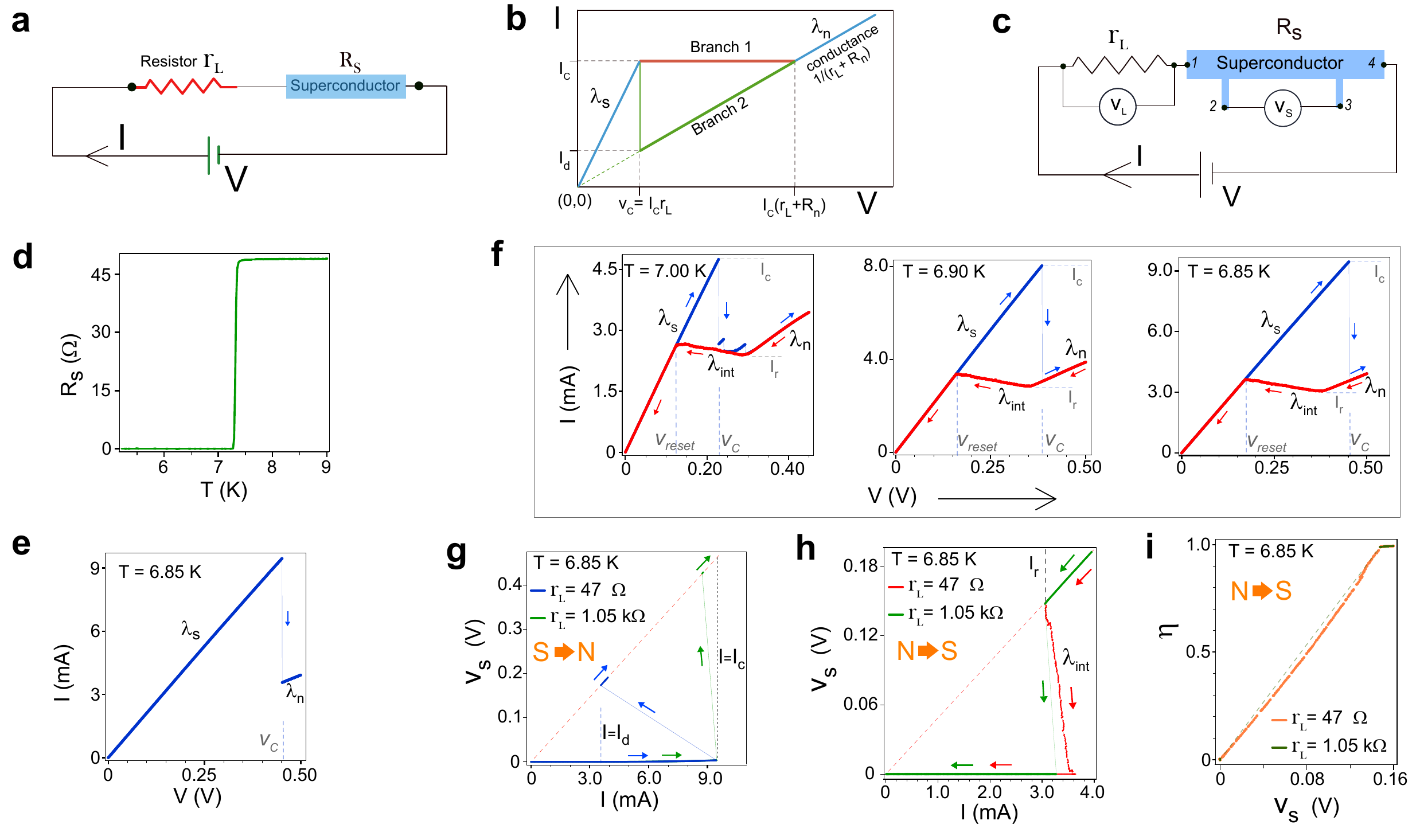}
\caption{\textbf{(a)} Circuit for the measurement of current-voltage relations. \textbf{(b)} Expected plot of current ($I$) in the circuit as a function of applied voltage $V$. \textbf{(c)} Circuit used for electrical measurements on a superconducting film in Hall bar geometry. \textbf{(d)} Measurement of resistance ($R$) as a function of temperature ($T$) in four-probe configuration showing the superconducting transition. \textbf{(e)} $I$-$V$ curve measured for the  circuit outlined in (c) for the superconductor-to-normal transition at $T$ = 6.85 K. \textbf{(f)} $I$-$V$ curves measured for both directions of voltage sweep at three different temperatures. \textbf{(g,h)} Voltage drop ($v_s$) across the superconductor ploted as a function of current ($I$) for two different values of resistance $r_L$ at $T$ = 6.85 K. \textbf{(i)} Plot of the parameter $\eta$ as a function of $v_s$ for the normal-to-superconductor transition.}
\end{center}
\end{figure*}

In his pioneering work on the thermodynamics of irreversible processes\cite{prigogine,glansdorff, glansdorff2, prigogine2} recognized by the Nobel Prize in Chemistry (1977), Prigogine along with his colleagues showed that non-equilibrium states are guided in many cases by the \emph{principle of minimum entropy production}. It states that when a system is not allowed to attain thermodynamic equilibrium by imposing certain  boundary conditions, it settles down to the state of least dissipation. There is a caveat that its validity is often limited only very close to equilibrium conditions. When translated to electrical circuit terminology\cite{landauer}, the principle remains valid for linear resistor networks where it reduces to Kirchhoff's current and voltage laws. We will now consider a different situation where the assumption of linearity is not valid. Let us take non-linear resistors realized with phase transition materials such that the relative contribution of two different phases appears as a new variable in the optimization problem. We may ask if the principle of minimum entropy production still holds true in such cases. This question will be addressed in this work for the specific case of superconductors.

We denote the resistance of a superconducting film as $R_s$ and its resistance in the normal state as $R_n$. The relation between these two quantities can be expressed as $R_s = \eta R_n$, where $\eta$ varies between 0 and 1 and represents the contribution of the normal phase to the total resistance. Transport properties of superconductors are generally probed under current-biased conditions due to their low resistance even in the normal state. At a temperature $T$ less than the critical temperature $T_c$, when the applied current ($I$) in a superconductor is increased it undergoes a transition from the superconducting ($\eta$=0) to the normal state ($\eta$=1) at a critical current $I_c$. The voltage across the system changes abruptly by an amount $I_cR_n$ at the critical current transition. The circuit to be used in our experiments is oulined in Fig. 1a. $r_L$ is a load resistor in series with the superconductor and $V$ is the voltage applied. The current in the circuit is given by

\begin{equation}
I=\frac{V}{r_L+\eta R_n}
\end{equation}

Small changes in the applied voltage $V$ can be adjusted by changes in both current and the parameter $\eta$.

\begin{equation}
\delta V = r_L \delta I + \eta R_n \delta I + IR_n  \delta \eta
\end{equation}

Measurements with current-biasing are done in the limit $r_L$$>>$$ R_n$, such that only the first term in the right-hand-side of Eq. (2) is important. We will conduct experiments in a very different condition where $r_L$ is either similar in magnitude or less than $R_n$. Thus, the term with $\delta \eta$ becomes relevant. We will call this mode of biasing as \emph{intermediate bias}, since it allows  us to measure current-carrying steady states in a wide range of conditions that do not impose either a fixed current or a fixed voltage on the superconductor. An important consequence of this intermediate biasing mode is that Eq. (2) involves two unknowns ($\delta I$ and $\delta \eta$) which does not lead to a unique solution. Arriving at a solution requires additional inputs from the physics of non-equilibrium conducting states that govern the variation of $\eta$ for different values of current and voltage across the superconductor. The nature of this dependence will be revealed in our measurements.

The current-voltage plot expected for the setup of Fig. 1a is described in Fig. 1b. At low voltage bias, $R_S$=0 and the entire voltage drops across $r_L$. This is marked as the branch $\lambda_s$, which continues till the voltage value $v_c$$=$$I_cr_L$ when the current reaches the critical current $I_c$ of the superconductor. Beyond this point ($V$$>$$v_c$), a voltage drop must develop across the superconductor. In the limit of large $V$, the superconductor is in the normal state ($R_S$$=$$R_n$). This is the branch $\lambda_n$ with conductance $\frac{1}{r_L+R_n}$. For intermediate voltages satisfying $v_c$$<$$V$$<$$I_c(r_L$$+$$R_n)$, a few possibilities we can guess are the following: (i) The system evolves along Branch 1, such that $\eta$ changes gradually from 0 ($\lambda_s$ line) to 1 ($\lambda_n$ line) with the current being approximately constant at $I_c$; (ii) the entire superconducting sample becomes normal as soon as a tiny voltage drop develops across it, evolving along Branch 2 which continues as the line $\lambda_n$; or (iii) any other trajectory between Branch 1 and 2. We denote $v_s$ as the voltage drop ($IR_s$) across the superconductor. If the superconductor is biased with a small current $I$, the power dissipation rate is $I^2R_s$. It is minimized for minimum resistance, corresponding to the zero-resistance superconducting state (provided $I$$<$$I_c$). On the other hand, for a small applied voltage $v_s$, the dissipation rate is $v_s^2/R_s$ which is minimized for minimum conductance (maximum resistance), driving the system towards a resistive state. Thus, the criterion of minimum dissipation predicts completely different steady states for the limits of current and voltage biasing.

Our experiments were conducted with a film of superconducting Nb patterned in the shape of a Hall bar. The width and thickness were 45 $\mu$m and 85 nm respectively. The circuit outlined in Fig. 1a was implemented by measuring the resistance ($R_S$) in four-probe configuration (Fig. 1c). The length of the Hall bar device (between the probes marked `1' and `4' for injecting current in Fig. 1c) was 960 $\mu$m. The distance between the voltage probes (`2' and `3') was 590 $\mu$m. The resistor $r_L$, measuring 47 $\Omega$, was at room temperature. The electrical lines inside the cryostat leading to the superconducting device had resistances less than 1 $\Omega$ each. The current in the circuit was determined by measuring the voltage $v_L$ across $r_L$. The $T_c$ of the superconducting film (Fig. 1d) was found to be 7.3 K, with a transition width less than 80 mK. From magnetotransport measurements, the Ginzburg-Landau coherence length ($\xi$) and electronic mean free path ($l$) were determined to be 7.3 nm and 2.0 nm respectively. This is a three-dimensional superconductor since its thickness is much larger than both $\xi$ and $l$.

We now present the experimentally observed $I$-$V$ relation (Fig. 1e) for the circuit in Fig. 1c, conducted by maintaining the superconductor at a temperature of 6.85 K (less than $T_c$). As the voltage bias $V$ in the circuit was continuously increased, the  critical current was reached at $v_c$ = 0.452 V. Following this, the current dropped sharply. It shows that as soon as a small voltage appeared across the superconductor, it transitioned immediately into the normal state. This extremely sharp feature of negative differential conductance (NDC) in the $I$-$V$ curve corresponds to Branch 2 in Fig. 1b - when the current $I$ drops to a value $I_d$ less than the critical current $I_c$. This confirms that a superconductor with an applied voltage difference follows the prediction of the principle of minimum entropy production by attaining the state of minimum conductance, which in this case is the normal state with $\eta$$=$$1$.

We will now discuss the details of the observed $I$-$V$ relations (Fig. 1f) at different temperatures below $T_c$ of the superconducting film. We first focus on the features of the transition along the direction of increasing $V$, i.e., along the superconductor-to-normal (\textbf{S}$\rightarrow$\textbf{N}) transition. Slightly below the critical temperature, at $T$ = 7.00 K, a large drop in conductance (reduction in current) is seen, although it does not drop sufficiently to reach the normal state on the line $\lambda_n$. We will discuss this specific case in more detail later. At 6.90 K and 6.85 K, the NDC phenomenon appears in one large step at $V$=$v_c$, reaching the normal state.

When the voltage $V$ is swept in the reverse direction of the normal-to-superconductor (\textbf{N}$\rightarrow$\textbf{S}) transition, the sharp NDC phenomenon is absent (Fig. 1f). The evolution of the $I$-$V$ curve from the branch $\lambda_n$ (when $R_s$=$R_n$) to the branch $\lambda_s$ (when $R_s$=0) occurs across an intermediate region (marked as $\lambda_{int}$) which leads to a hysteresis loop. The transition happens gradually as $V$ is ramped down, and is complete at a voltage $v_{reset}$ when $R_s$ becomes zero again. There are two possible mechanisms that are considered to be important for hysteresis in superconductors. One mechanism is the interplay of two different timescales\cite{ivlev,vodolazov}: (i) $\tau_\phi$, which is the relaxation time of the order parameter phase and varies typically as $\tau_\phi$ $\sim$ $1/I$, and (ii) $\tau_\Delta$, which is the relaxation time of the order parameter amplitude and depends on the inelastic relaxation time. The critical current transition marks a departure from the equilibrium state of zero resistance. Usually $\tau_\phi$$<<$$\tau_\Delta$ after the normal state appears at $I \simeq I_c$, and this prevents the superconducting state to form back even if the current is ramped down from $I_c$. It has to reduce sufficiently to a lower value $I_r$ (retrapping current) to allow the reappearance of the superconducting state. The second mechanism for hysteresis is a Joule heating effect\cite{markovic}. When the superconducting state is destroyed by application of a current larger than $I_c$, the generation of heat due to the finite resistance causes the temperature of the system to increase beyond $T_c$. Superconductivity can reappear only when the current is reduced to a lower value $I_r$, such that the dissipated heat is incapable of maintaining a high temperature. Either one of these two mechanisms may be responsible for the hysteresis seen in our measurements. In Fig. 1f, the branch $\lambda_{int}$ shows the evolution of the system for the \textbf{N}$\rightarrow$\textbf{S} transition towards the zero resistance state with $I$ $\simeq$ $I_r$.

The magnitude of the current $I_d$ (as shown in Fig. 1b) is related to $I_c$ following the relation:
\begin{equation}
I_d=\frac{I_c}{1+\frac{R_n}{r_L}}
\end{equation}

If the external resistance $r_L$ is made smaller, $I_d$ is supposed to reduce as well. In Fig. 1f, for temperatures of 6.90 K and 6.85 K, the conductance switching at $V$=$v_c$ results in a lower current $I_d$ that directly reaches the normal state on the $\lambda_n$ line. The retrapping current $I_r$ (on the line $\lambda_{int}$) is smaller than $I_d$. The gap between $I_c$ and $I_r$ of superconductors generally reduces as the temperature increases towards $T_c$. This leads to a different situation where $I_r$ becomes larger than the predicted $I_d$ (for a given value of $r_L$) closer to $T_c$, as we see for $T$ = 7.00 K in Fig. 1f. In this case, if at $V$=$v_c$ the current has to drop to the expected value $I_d$ on the $\lambda_n$ line, then it needs to reduce further than the retrapping current $I_r$. We observe that in reality this does not happen, and the minimum value of current for the \textbf{S}$\rightarrow$\textbf{N} sweep direction coincides with the retrapping current for the reverse sweep direction (\textbf{N}$\rightarrow$\textbf{S}) indicated by the line $\lambda_{int}$.

As a hypothetical case, let us assume $r_L$ to be extremely small in the limit $r_L\rightarrow0$. It is then expected from Eq. 3 that $I_d\rightarrow0$. This would imply that the superconductor becomes normal as soon as a finite voltage difference develops for $V$$>$$v_c$ carrying an infinitesimally small current, thus reaching a state almost at thermodynamic equilibrium. However, this assertion runs into a contradiction with the condition of \textit{minimum free energy} applicable to equilibrium states which determines the system to be superconducting for $T$$<$$T_c$. Therefore, it is not possible to reach an arbitrarily small value of $I_d$ on physical grounds and there must exist certain restrictions on how low the current can reduce at $V$=$v_c$. From our experiments, we see that the fundamental limit is set by the retrapping current $I_r$. As a general rule, we observe that the conductance switching behaviour leads to a reduction of the current to the predicted $I_d$ (given by Eq. 3) when $I_r$$<$$I_d$, but when $I_r$$>$$I_d$ it stays slightly above $I_r$. While the principle of minimum entropy production remains valid since the conductance of the superconductor reduces sharply for $V$$>$$v_c$, it is also important to note that the resulting state follows the limitations set by the general thermodynamic properties of superconductors.

Till now, we have presented the results of $I$-$V$ measurements using a low resistance $r_L$, which is very far from the case of current-biasing. We will now compare these results with more conventional current-biased measurements. Current-biasing was done using the same setup in Fig. 1c, by choosing a much larger load resistance ($r_L>>R_n$) and adjusting the source voltage $V$ appropriately. We used $r_L$ = 1.05 k$\Omega$. Fig. 1g shows the voltage drop ($v_s$) across the superconductor as a function of current $I$ for $T$ = 6.85 K, for the \textbf{S}$\rightarrow$\textbf{N} transition. At the critical current transition ($I \approx I_c$), the film becomes normal and $v_s$ becomes finite as resistance develops. The small negative slope in the $v_s$-$I$ plot for $r_L$ = 1.05 k$\Omega$ arises due to the finite $R_n/r_L$ ratio. We compare these results with the variation of $v_s$ as a function of $I$, observed in the $I$-$V$ relations earlier using $r_L$ = 47 $\Omega$ (shown in Fig. 1f). In that case, as soon as a resistance developed in the normal state (\textbf{S}$\rightarrow$\textbf{N} transition), the current reduced abruptly from $I_c$ to the lower value $I_d$. A consequence of this fact is that, for the case $r_L$ = 47 $\Omega$, the function $v_s$($I$) is bistable for the range  $I_d$$<$$I$$<$$I_c$ (Fig. 1g). In the intermediate biasing method described here, lowering the value of $r_L$ results in lowering the value of $I_d$ (Eq. 3), which in turn leads to the observation of the bistable region over a larger current range in the $v_s$-$I$ scatter plot. This shows a clear distinction of the voltage-biased limit (low $r_L$) from the current-biased case (large $r_L$). For the latter, the normal state appears for $I$$>$$I_c$ and the $v_s$-$I$ relation is  single-valued.

Bistability has been observed previously in superconducting devices, owing to the occurrence of phase slip centres in nanowires in the quasi-one-dimensional limit\cite{michotte} and due to thermal hotspots in narrow microbridges\cite{skocpol}. Such features have been theoretically predicted\cite{keizer} in voltage-biased short mesoscopic wires due to a non-thermal energy distribution of quasiparticles. In these cases, bistable $I$-$V$ curves were associated with  properties specific to low dimensional systems and finite size effects. In contrast, the phenomenon reported in this work is much more general as it occurs in macroscopic samples. The three-dimensional nature of the superconducting Nb film is evident from the sharpness of the critical current transition (Fig. 1g). In our case, there is no prerequisite of specific sample geometry or reduced dimensionality for the observation of such bistability. The feature of negative differential conductance arises here from the tendency of a superconductor to minimize dissipation under an applied voltage difference.

The $v_s$-$I$ plots for the reverse direction of normal-to-superconductor (\textbf{N}$\rightarrow$\textbf{S}) transition is shown in Fig. 1h. For $r_L$ = 1.05 k$\Omega$ (current-biasing limit), the voltage drop reduced abruptly to zero as the retrapping current $I_r$ is reached. An intriguing feature is observed when we compare this result with the case for $r_L$ = 47 $\Omega$ (which refers to the measurements  shown in Fig. 1f). For the latter, the transition is not abrupt and there are several data points on the $v_s$-$I$ scatter plot leading towards the zero resistance state. In Fig. 1i, we show the estimated value of $\eta$ (the ratio of $R_s$ measured and the resistance $R_n$ of the completely normal state) as a function of $v_s$ for the \textbf{N}$\rightarrow$\textbf{S} transition. The data points in the scatter plot for the case $r_L$ = 47 $\Omega$ show fractional values of $\eta$ between 1 and 0, which signifies the presence of both superconducting and normal fractions within the system. Coexistence of superconducting and normal phases is a well-known phenomenon in the mixed state of type-II superconductors in the presence of a magnetic field when flux penetration takes place in the form of vortices\cite{abrikosov,tinkham}. In our case, we observe that steady states bearing signatures of both phases can be generated even without an applied magnetic field by letting a voltage gradient develop across the superconductor. This highlights a unique aspect of the intermediate biasing method and the fact that it can reveal the existence of non-equilibrium conducting states that are inaccessible in current-biased measurements. These states correspond to the range $0$$<$$v_s$$<$$I_rR_n$, which allows the system to be neither completely normal nor completely superconducting. Since $\delta I$ is very small along the line $\lambda_{int}$ in Fig. 1h, the evolution of the system following Eq. 2 reduces to the expression: $\delta \eta \approx \frac{\delta V}{I_rR_n} $. Small changes in the applied voltage $V$ are adjusted by changes in the value of $\eta$.

We show in Fig. 2a the scatter plot of $v_s$ as a function of $I$ for $T=$7.00 K, which corresponds to the same measurement in Fig. 1f at the given temperature (with $r_L$= 47 $\Omega$) for the \textbf{S}$\rightarrow$\textbf{N} transition. Since the reduced current $I_d$ can not be attained at this temperature, we observe values of $v_s$ in-between the superconducting and normal states in Fig. 2a (marked by a dashed ellipse). This also implies fractional values of $\eta$ less than 1 (Fig. 2b) and the occurrence of some form of phase coexistence. These results are compared with nearly current-biased experiments using $r_L$= 1.05 k$\Omega$ (Figs. 2a and 2b). In this case, no steady state with fractional $\eta$ is observed. The general criterion for observing nonequilibrium phase coexistence during the superconducting-to-normal transition is that $I_d$ calculated from Eq. 3 must be lesser than $I_r$. This is possible either by approaching temperatures close to $T_c$ (when $I_r$ is closer to $I_c$), or by lowering the value of $r_L$ (which reduces the predicted $I_d$).

\begin{figure}
\begin{center}
\includegraphics[width=78mm]{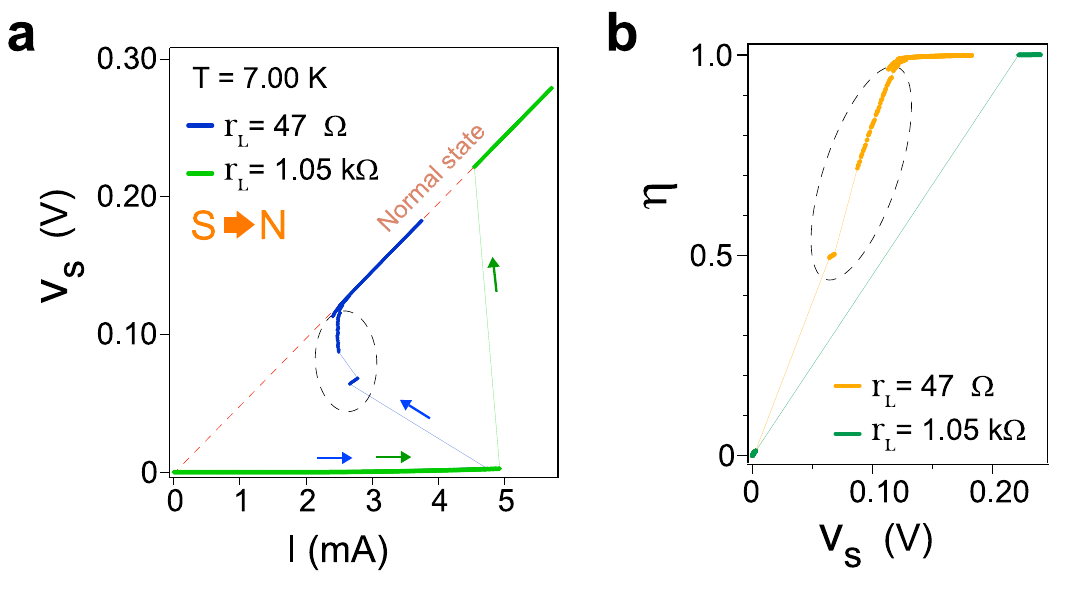}
\caption{\textbf{(a)} Voltage drop ($v_s$) across the superconductor plotted as a function of current ($I$) at $T$ = 7.00 K. This shows the superconductor-to-normal transition for two different measurements using different values of $r_L$. The states within the dashed ellipse correspond to the nonequilibrium  phase coexistence phenomenon. \textbf{(b)} Plot of the parameter $\eta$ as a function of $v_s$ for the superconductor-to-normal transition.}
\end{center}
\end{figure}

Our results described above underscore the possibility that there may exist a deep connection between transport phenomena in superconductors and the general framework of non-equilibrium statistical mechanics as expounded by Prigogine. A central concept emphasized by Prigogine was that irreversible processes may lead to an organization of matter into \emph{dissipative structures}, which do not exist under conditions of thermodynamic equilibrium\cite{prigogine}. These ideas have been quite influential\cite{glansdorff2,prigogine2} in different problems of physical chemistry and hydrodynamics. The states with fractional $\eta$ in Fig. 1i (for $r_L$ = 47 $\Omega$) represent a particular form of ordering of the superconducting and normal fractions in a non-equilibrium steady state concomitant with a finite rate of dissipation. This provides an example of a dissipative structure in a superconductor. An interesting problem here is the question of what type of ordering principle it represents. There are two possibilities. The first one is that there is a physical segregation of the system into superconducting and normal regions along its length, with the latter occupying a fraction $\eta$ of the total length. The second possibility is that the superfluid density fluctuates in time, such that the charge carriers oscillate between the superconducting and normal phases. In this case, $\eta$ is a more complex quantity signifying the contribution of the normal fraction to dissipative mechanisms in transport. This latter model is similar to the description of charge transport phenomena with respect to phase slips and weak links in superconductors\cite{ivlev}, with the additional requirement that in our case it must be adapted to large macroscopic samples. Further research will be required to understand in more detail the structure and distribution of electronic phases in this transport regime.

In summary, we have conducted current-voltage measurements in large superconducting films without imposing the constraint of constant current bias. These experiments reveal a sharp feature of negative differential conductance, which shows that the principle of minimum entropy production holds true for the transition from the zero resistance superconducting state to the normal state. For the reverse normal-to-superconductor transition in the vicinity of the retrapping current, our measurements reveal the features of a state that has resistance intermediate between the two phases and indicates some form of admixture of both. This type of nonequilibrium phase coexistence is also visible in some specific instances during the superconductor-to-normal transition. Our results were obtained on large three dimensional films, showing that this physics does not originate from effects of finite size or reduced dimensionality. This work demonstrates how novel varieties of nonequilibrium steady states can be engineered in superconductors by carefully changing the biasing conditions. Understanding current-voltage relations of superconductors is an  extremely important topic for applications in superconducting electronics, which has generated a lot of interest at present in the context of superconducting diodes\cite{hu,ando,hou,lin,pal}. The experimental method demonstrated here has the potential to inspire novel approaches towards the investigation of such problems.

\section{Acknowledgments}
The author thanks Swastik Bhattacharya, Claire Marrache-Kikuchi, Miguel Monteverde, Mark Goerbig, Marc Gabay, Mandar Deshmukh and Aviad Frydman for insightful discussions, and Laurent Berg\'{e} and Sophie Gu\'{e}ron for help with fabrication of samples.



\end{document}